\documentstyle[amssymb,aps,prl,floats,graphicx]{revtex}

\begin{document}

\twocolumn[\hsize\textwidth\columnwidth\hsize\csname
@twocolumnfalse\endcsname

\title{Shear flow and Kelvin-Helmholtz instability in superfluids}

\author{
    R.~Blaauwgeers$^{1,2}$, V.~B.~Eltsov$^{1,3}$,
    G.~Eska$^{1,4}$, A.~P.~Finne$^{1}$, R.~P.~Haley$^{1,5}$,
    M.~Krusius$^{1}$, J.~J.~Ruohio$^{1}$,\\ L.~Skrbek$^{1,6}$,
    and G.~E.~Volovik$^{1,7}$
}

\address{
  $^{1}$Low Temperature Laboratory,
  Helsinki University of Technology,
  P.O.Box 2200, FIN-02015
  HUT, Finland\\
  $^{2}$Kamerlingh Onnes Laboratory, Leiden University, P.O.Box
  9504, 2300 RA Leiden, The Netherlands\\
  $^{3}$Kapitza Institute for Physical Problems, Kosygina 2, 117334
  Moscow,  Russia\\
  $^{4}$Physikalisches Institut, Universit\"at Bayreuth, D-95440
  Bayreuth, Germany\\
  $^{5}$Department of Physics, Lancaster University, Lancaster, LA1\,4YB,
  UK\\
  $^{6}$Joint Low Temperature Laboratory, Institute of Physics ASCR and Charles University, V
  Hole\v{s}ovi\v{c}k\'ach 2,\\ 180\,00 Prague, Czech Republic\\
  $^{7}$Landau Institute for Theoretical Physics, Kosygina 2, 117334
  Moscow, Russia
}


\date{\today}
\maketitle

\begin{abstract}

The first realization of instabilities in the shear flow between
two superfluids is examined. The interface separating the A and B
phases of superfluid $^3$He is magnetically stabilized. With
uniform rotation we create a state with discontinuous tangential
velocities at the interface, supported by the difference in
quantized vorticity in the two phases. This state remains stable
and nondissipative to high relative velocities, but finally
undergoes an instability when an interfacial mode is excited and
some vortices cross the phase boundary. The measured properties of
the instability are consistent with a modified Kelvin-Helmholtz
theory.
\end{abstract}
]
\narrowtext

Instabilities in the shear flow between two layers of fluids
\cite{HydroDynamics} belong to a class of interfacial
hydrodynamics which is attributed to many natural phenomena.
Examples are wave generation by wind blowing over water
\cite{LordKelvin}, the flapping of a sail or flag in the wind
\cite{Rayleigh,FlexibleFilament}, and even flow in granular beds
\cite{Granular}. In the hydrodynamics of inviscid and
incompressible fluids the transition from calm to wavy interfaces
is known as the Kelvin-Helmholtz (KH) instability
\cite{Helmholtz,LordKelvin}. Since Lord Kelvin's treatise in 1871,
difficulties have plagued its description in ordinary fluids,
which are viscous and dissipative. They also display a shear-flow
instability, but its correspondence with that in the ideal limit
is not straightforward. The tangential velocity discontinuity in
the shear-flow instability is created by a vortex sheet. In a
viscous fluid a planar vortex sheet is not a stable equilibrium
state and not a solution of the hydrodynamic equations
\cite{Birkhoff}.

Superfluids provide a close variation of the ideal inviscid limit
considered by Lord Kelvin and thus an environment where the KH
theory can be tested. The initial state is a non-dissipative
vortex sheet -- the interface between two superfluids brought into
a state of relative shear flow. So far the only experimentally
accessible case where this can be studied in stationary
conditions, is the interface between $^3$He-A and $^3$He-B
\cite{VollhardtWoefle}, where the order parameter changes symmetry
and magnitude, but is continuous on the scale of the superfluid
coherence length $\xi \sim 10\,$nm. We discuss an experiment,
where the two phases slide with respect to each other in a
rotating cryostat: $^3$He-A performs solid-body-like rotation
while $^3$He-B is in the vortex-free state and thus stationary in
the laboratory frame. While increasing the rotation velocity
$\Omega$, we record the events when the AB phase boundary becomes
unstable -- when some circulation from the A-phase crosses the AB
interface and vortex lines are introduced into the initially
vortex-free B phase. On increasing the rotation further, the
instability occurs repeatedly. Such a succession of instability
events can be understood as a spin-up of $^3$He-B by rotating
$^3$He-A.

\begin{figure}[t]
\centerline{\includegraphics[width=0.9\linewidth]{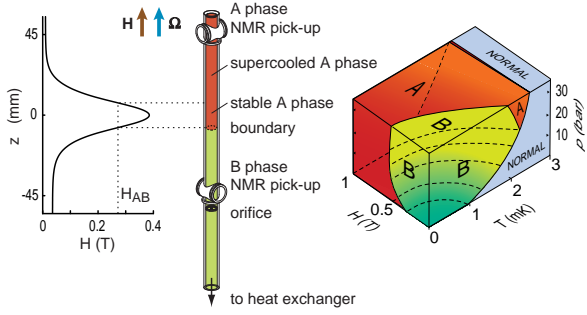}}
\medskip
\caption{ Stabilization of the first-order $^3$He-A -- $^3$He-B
phase boundary. At pressures $p \geq 21\,$bar, A-phase extends to
lower temperatures in external magnetic field (see phase diagram
on the right). The sample (length 11\,cm, radius $R = 0.3\,$cm) is
first cooled to A phase. On further cooling, the A$\rightarrow$B
transition happens in the coldest place at the bottom. The AB
boundary then starts to move up and rises to a height $z$ where
the barrier field $H(z)$ equals the value of the thermodynamic
A$\rightarrow$B transition $H_{\rm AB}(T,p)$.  The A phase in the
top section remains in a metastable supercooled state
\protect\cite{Schiffer}. Ultimately, the boundary disappears when
$H_{\rm AB}(T,p) > [H(z)]_{\rm max}$. For $p = 29.0\,$bar and
$H(z)$ as shown in the figure with a current of 4\,A in the
barrier solenoid, the stable AB boundary exists below $ 2.07\,$mK
down to $ 1.33\,$mK. The NMR spectrometers operate in homogeneous
static magnetic fields of 10\,mT and 35\,mT, chosen for best
measuring sensitivity of the single-vortex signal.}
\label{AB_Sample}
\end{figure}

Our experimental setup is shown in Fig.~\ref{AB_Sample}. The AB
boundary is forced against a magnetic barrier in a smooth-walled
quartz container, by cooling the sample below $T_{\rm AB}$ at
constant pressure in a rotating refrigerator. The number of
vortices in both phases is independently determined from the
simultaneously measured nuclear magnetic resonance (NMR) spectra
\cite{Parts,DoubleQuantumVortex}. By tuning the barrier field or
the temperature, the state of the sample can be changed from all A
phase to all B phase or to a two-phase configuration. The
evolution of the quantized vorticity as a function of $\Omega$ is
then observed to be radically different when the AB interface is
present.

The quasi-isotropic $^3$He-B supports singly quantized vortices
with a core size comparable to $\xi$ \cite{Parts}. In the
anisotropic $^3$He-A we form vortex lines
\cite{DoubleQuantumVortex} with continuous skyrmion topology:
Inside its central part the order-parameter amplitude remains
constant but the axis of the orbital anisotropy covers a solid
angle of $4\pi$. Such a ``soft-core'' structure carries continuous
vorticity with two circulation quanta and is three orders of
magnitude larger than the core of the B-phase vortex. Thus
converting an A-phase vortex into a B-phase vortex requires large
concentration of the flow energy.

\begin{figure}[t]
\centerline{\includegraphics[width=0.9\linewidth]{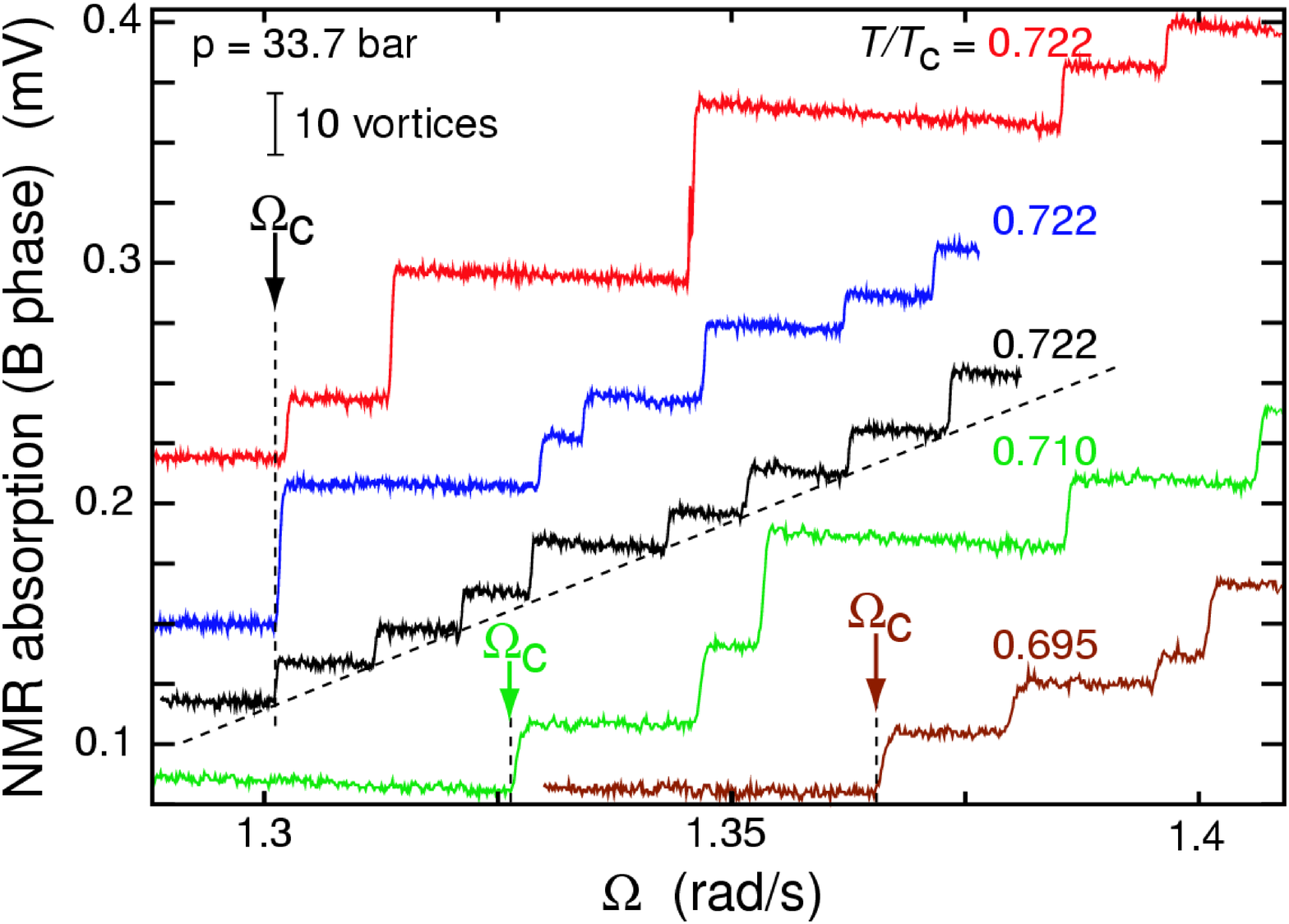}}
\medskip
\caption{Instability events during slowly increasing rotation
($d\Omega /dt = 2.5 \cdot 10^{-4}$ rad/s$^2$). NMR absorption
signals as a function of $\Omega$ are recorded (shifted
arbitrarily on the vertical scale). The vertical jumps mark events
in which $\delta N$ new vortex lines enter the B phase section of
the sample. The height of each jump is proportional to $\delta N$,
a small stochastic number. When averaged over a large interval of
$\Omega$, the number of B-phase vortex lines grows linearly with
$\Omega$. This is demonstrated also by the dashed line: the
instability occurs independently of $\Omega$ at constant critical
drive, {\it ie.} $u_{\rm c} \! = \! |v_{\rm sB}- v_{\rm n}|_{r=R}
\! = \! \Omega_{\rm c} R \! = \, $const. The three top most signal
traces were recorded at the same temperature, ie. measurements at
fixed $T$ yield the same $\Omega_{\rm c}$ (if small variations in
$v_{\rm cA}$ are accounted for \protect\cite{CritVelA}). The two
bottom traces at different $T$ illustrate that $\Omega_{\rm c}$
depends on temperature. Here $T_{\rm AB}(H=0)= 0.785\,T_{\rm c}$.
} \label{InstabilitySteps}
\end{figure}

The large difference in core radii is the origin for the much
lower rotation at which vortices start forming in A phase at the
outer sample circumference, compared to B phase \cite{Parts}. If
the sample consists of only B phase then the critical velocity for
forming the first vortex line in the setup of Fig.~\ref{AB_Sample}
is $v_{\rm cB} > 7\,$mm/s \cite{CritVelB}, while the critical
velocity $v_{\rm cA}$ is a factor of 20 smaller \cite{CritVelA},
independently of the presence of the AB boundary.

Vortex lines are thus easily created at low rotation in the
A-phase section of the sample, while no vortices are detected in
the B phase section. This is the ideal non-dissipative initial
state where the two superfluid phases slide along each other
without friction. When $\Omega$ is increased further, sudden
bursts of vortex lines are observed in the B-phase section, as
shown in Fig.~\ref{InstabilitySteps}. The onset $\Omega_{\rm c}$
depends on temperature and on the current in the barrier magnet.
Overall, the measured characteristics of the bursts fit a shear
flow instability, which provides a mechanism for the circulation
to cross the AB interface.

\begin{figure}[t]
\centerline{\includegraphics[width=0.9\linewidth]{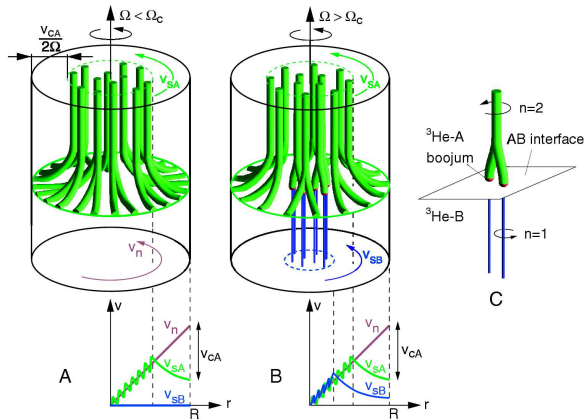}}
\medskip
\caption{The two configurations of vortex lines with the AB
boundary in rotation. ({\bf A}) $\Omega< \Omega_{\rm c}$: Vortex
lines are formed in $^3$He-A while $^3$He-B remains vortex-free.
Near the AB boundary the A-phase vortices bend parallel to the
interface and form a vortex sheet between the sliding superfluids.
The radial distributions of the normal and superfluid velocities
far from the AB interface are depicted below, with $v_{\rm
nA}=v_{\rm nB}=\Omega r $, while $v_{\rm sB}=0$. ({\bf B}) $\Omega
\geq \Omega_{\rm c}$: Vortices are observed to appear in the B
phase in events of a few lines at a time. They form a central
cluster in the B-phase section. ({\bf C}) A hydrodynamically
stable state with respect to externally imposed perturbations in
$\Omega$, $T$, or $H$ exists at the AB interface for $\Omega \geq
\Omega_{\rm c}$.  A topologically stable configuration for the
vortex-line intersection with the AB interface is suggested in
Ref. \protect\cite{Volovik}: The doubly quantized A-phase vortex
terminates at the AB interface in two point singularities, known
as boojums. These, in turn, are the end points of two singly
quantized B-phase vortices. } \label{AB_Configuration}
\end{figure}

$^3$He-A and $^3$He-B are states of the same order-parameter
manifold. One of the conditions on their phase boundary is that
the phase of the order parameter has to be continuous
\cite{Volovik}: If the circulation is not continuing across the
interface ( Fig.~\ref{AB_Configuration}A) then the existing
A-phase vortex lines have to bend and form a vortex layer on the
AB interface. This is the only hydrodynamically stable state, with
vortex-free flow seen by the B-phase spectrometer and a large
cluster of vortex lines detected by the A-phase spectrometer. The
build up of a sheet-like vortex layer means that the A-phase
circulation cannot easily penetrate into the vortex-free B-phase
and the AB interface remains stable up to the measured critical
velocity $u_{c} = \Omega_{\rm c} R \sim 2$ -- $4\,{\rm mm/s} <
v_{\rm cB}$.

At $\Omega > \Omega_{\rm c}$, some vortex lines have broken
through the AB phase boundary. Thus there exists also a stable
configuration in which vortices from the A phase continue into the
B phase, where they form a cluster in the center
(Fig.~\ref{AB_Configuration}B). The likely topology of the
intersection is illustrated in Fig.~\ref{AB_Configuration}C. Thus,
although the AB interface is not directly monitored by the two NMR
spectrometers, the hydrodynamically stable states below and above
$\Omega_{\rm c}$ can only be the configurations in
Fig.~\ref{AB_Configuration}. By comparing B-phase vortex-line
creation in the presence and absence of the AB interface, we
conclude that the events in Fig.~\ref{InstabilitySteps} originate
from the AB phase boundary.

\begin{figure}[t]
\centerline{\includegraphics[width=0.9\linewidth]{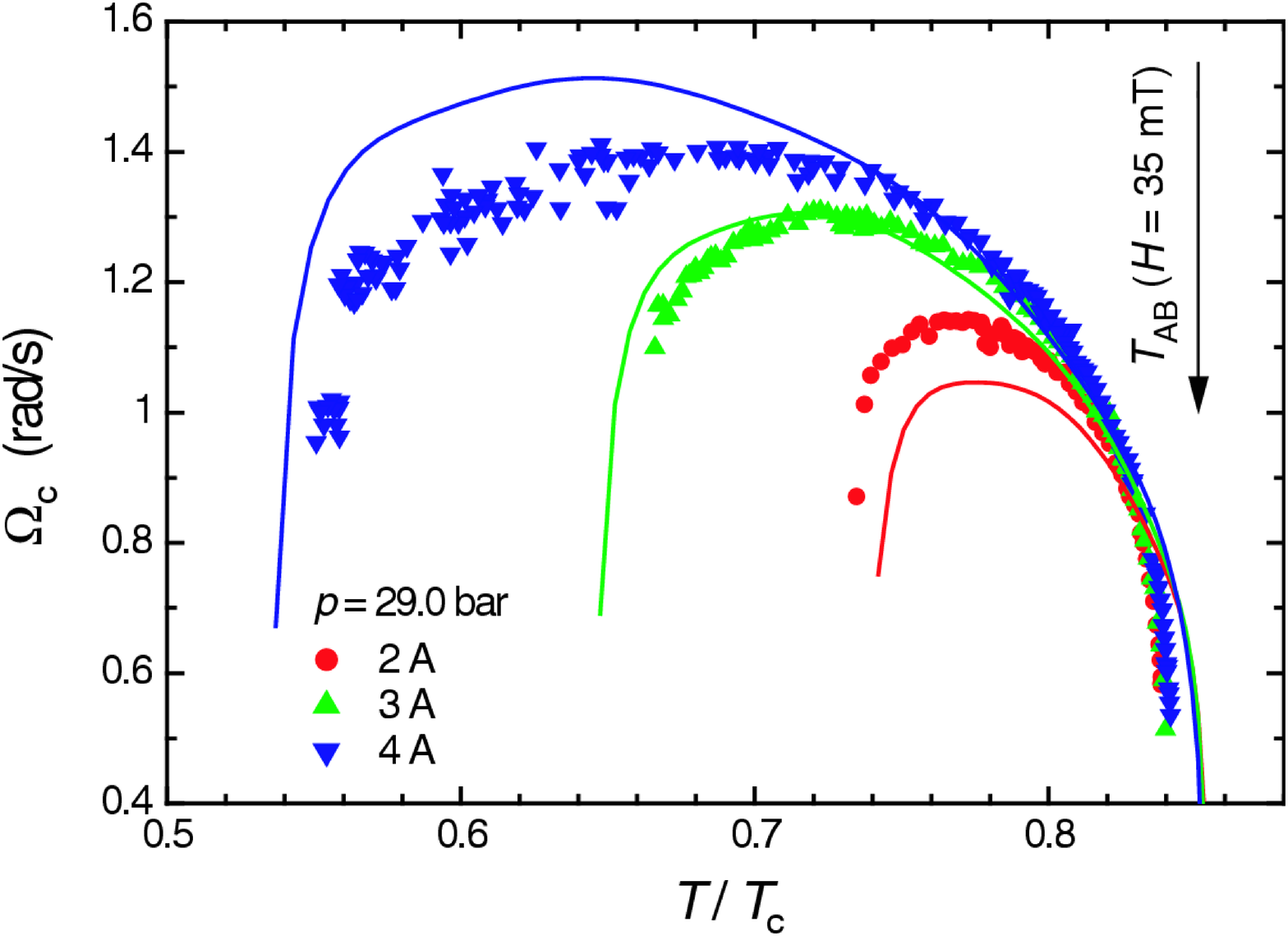}}
\medskip
\caption{Critical velocity $\Omega_{\rm c}$ for the first
appearance of B-phase vortex lines in the presence of the AB
boundary as a function of temperature, while $\Omega$ is slowly
increased ($\dot\Omega = 5 \cdot 10^{-4}\,$rad/s$^2$). The current
in the barrier solenoid is constant but different for the three
sets of data. It controls the magnetic force $F$ which is largely
responsible for the shape of the curves. The solid curves
represent Eq.~(\protect\ref{Anisotropic6}) if one sets
$\Omega_{\rm c} = |v_{\rm sB}- v_{\rm n}|/R$ and $v_{\rm sA} =
v_{\rm n}$. No fitting parameters are used. The values for
$\sigma(T)$, $\chi_A-\chi_B(T,H)$, and $\rho_{\rm s}(T)$ are
obtained from accepted references
\protect\cite{VollhardtWoefle,Osheroff,Wheatley,Bozler}, while $H$
and $\nabla H$ apply for the profile $H(z)$ at the location of the
AB boundary: $H_{\rm AB}(T) = H(z)$. } \label{KH_Onset}
\end{figure}

In the classical KH instability the interface between fluids with
densities  $\rho_1$ and $\rho_2$ becomes destabilized by inertial
effects, which are normally balanced by gravity $g$ and surface
tension $\sigma$. The relative velocity $|v_2-v_1|$ acts as a
drive. When it reaches a critical value \cite{LordKelvin} given by
\begin{equation}
 {\rho_1 \rho_2 \over
\rho_1+\rho_2} (v_2-v_1)^2=2\sqrt{\sigma F},
\label{InstabilityCondition}
\end{equation}
where $F = g (\rho_1 -\rho_2)$ is the gravitational restoring
force, waves with wave vector $k=\sqrt{F/\sigma}$ are created on
the interface.

This approach can be generalized for superfluids in terms of
two-fluid hydrodynamics \cite{Tilley}. For superfluid $^3$He under
our experimental conditions it is safe to assume that the normal
fractions are always in solid-body rotation. Instead of gravity,
the restoring force is now produced by the magnetic barrier
$H(z)$, owing to the difference in the susceptibilities $\chi_{\rm
A}$ and $\chi_{\rm B}(T,H)$: $F = (1/2)(\chi_{\rm A} -\chi_{\rm
B}(T,H))\nabla (H^2)$ at $H_{\rm AB}(T)$. The restoring force can
thus be calculated as a function of temperature and current in the
barrier solenoid. The motion of the AB interface with respect to
the normal component is subject to a finite damping
\cite{Buchanan}. Since the initial state is non-dissipative,
damping does not explicitly appear in the stability condition, but
modifies the build-up rate of the interface perturbations when the
instability develops. The onset of the instability can be derived
from the dynamics of small amplitude perturbations, as
Eq.~(\ref{InstabilityCondition}) is generally derived in textbooks
\cite{HydroDynamics}, or from the thermodynamics when
perturbations of the interface lead to a negative free energy in
the rotating frame. The mode for which the interface first becomes
unstable has the same wave vector $k=\sqrt{F/\sigma}$ as before at
a drive given by
\begin{equation}
{1\over 2}\rho_{\rm sA}~ (v_{\rm sA}-v_{\rm n}) ^2 +{1\over
2}\rho_{\rm sB}~ (v_{\rm sB}-v_{\rm n}) ^2
    =  \sqrt{\sigma F}~,
\label{Anisotropic6}
\end{equation}
where $\rho_{\rm s}$, $\rho_{\rm n}$, and  $v_{\rm s}$, $v_{\rm
n}$ are the densities and velocities of superfluid and normal
components \cite{GEV}. In Fig.~\ref{KH_Onset} we plot the
prediction for the instability for three magnetic barrier
profiles. With no fitting parameters the agreement with experiment
is surprisingly good.

From Fig.~\ref{InstabilitySteps} it is seen that of order $\delta
N \! \approx \! 10$ vortex lines enter the B phase in one
instability: in fact, at $0.77\,T_{\rm c}$ (with $u_c
\!=\!0.39$\,cm/s at 29.0\,bar) the measured average for more than
100 events is $\delta N \! \approx \!11$. To compare with
Eq.~(\protect\ref{Anisotropic6}), $\delta N$ corresponds to the
number of circulation quanta $\kappa \! = \! h/2m_3$, which fit in
one corrugation of the interface mode of size $\lambda_{\rm c}/2
\!=\! \pi/k$. In practice, the counterflow  $|v_{\rm sA}-v_{\rm
n}|$ in A-phase is limited by the small critical velocity $v_{\rm
cA}$ \cite{CritVelA}, and we may set $v_{\rm sA}-v_{\rm n} \!
\approx \! 0$ in Eq.~(\protect\ref{Anisotropic6}). In solid-body
rotation there are then $N \!\approx \!\pi R^2 \, \Omega_{\rm
c}/\kappa$ vortex lines in the A phase which all flare out into
the lateral sample boundary at the AB interface
(Fig.~\ref{AB_Configuration}A). Measured along the perimeter of
the sample there are $\Omega_{\rm c}R/\kappa$ circulation quanta
per unit length and thus in one corrugation $\delta N \! \approx
\! ({\pi u_{\rm c}})/({k\kappa})$. From
Eq.~(\protect\ref{Anisotropic6}) this is seen to be $\delta N \!
\approx \! (2\pi \sigma)/(\kappa u_{\rm c}\rho_{\rm sB})$, which
at $0.77\,T_{\rm c}$ gives 9 vortex lines in agreement with the
measured number.

Eq.~(\protect\ref{Anisotropic6}) has more interesting properties.
As the normal components of both phases are in solid-body rotation
and thus at rest in the rotating frame, it is the ``superfluid
winds'' -- the flow of the superfluid component, $|v_{\rm
s}-v_{\rm n}|$, on each side of the interface -- which produce the
instability. It takes place even if the two superfluids have the
same densities and velocities. In this sense it resembles the
flapping flag instability discussed by Rayleigh \cite{Rayleigh}:
The role of the flagpole, which fixes the reference frame, falls
here on the normal component, which moves with the rotating
container.

It is interesting to compare Eqs.~(\ref{InstabilityCondition}) and
(\protect\ref{Anisotropic6}) in the low temperature limit, when
$\rho_{\rm n} \rightarrow 0$ and $\rho_{\rm sA} \approx \rho_{\rm
sB}\rightarrow \rho$. We find that
Eq.~(\protect\ref{Anisotropic6}) gives a critical velocity which
is by $\sqrt{2}$ smaller than that from
Eq.~(\ref{InstabilityCondition}). In fact, at $T=0$ the valid
condition is Eq.~(\ref{InstabilityCondition}), since $\rho_{\rm
n}$ is exactly zero and no normal component is left to provide a
reference frame for the now undamped interface \cite{GEV}. The
ideal KH condition is then restored, as the only remaining drive
is the velocity difference between the superfluid components. Not
surprisingly, the two cases $\rho_{\rm n}=0$ and $\rho_{\rm
n}\rightarrow 0$ give different onsets for the instability: For a
finite-size system the result would depend on the observation time
-- the time one waits for the interface to be coupled to the
laboratory frame.

The shear-flow instability in superfluids thus provides an
explicit bridge between the ideal inviscid and the viscous
hydrodynamics. For two-fluid hydrodynamics it may have important
consequences. It represents a new surface mechanism for vortex
generation which might be applicable also to other interfaces, of
which the normal state -- superfluid boundary is of major
interest. This interface occurs, for example, when the bulk
superflow instability limit is reached in rotating $^3$He-B at the
sharpest surface spike on the lateral sample boundary
\cite{Parts}. A bubble of normal liquid is then created around
this spike. Since the surrounding superflow moves at a high
relative velocity, the bubble's interface may undergo a shear-flow
instability and promote the generation of a vortex line.

A related case has been investigated in bulk $^3$He-B which is in
vortex-free rotation while it is irradiated with slow neutrons
\cite{Ruutu}. The absorption process of one neutron creates
locally a bubble of $100\,\mu$m size which is overheated up to the
normal state. The bubble then quench cools back to the temperature
of the surrounding superflow, as the normal--superfluid interface
of the bubble contracts at high velocity. The measurements prove
directly that vortices are formed in this rapid non-equilibrium
transition and that the process is consistent with a volume effect
known as the cosmological Kibble-Zurek (KZ) mechanism
\cite{Kibble,Zurek}. It has been suggested as the origin for
cosmic-string formation when rapid non-equilibrium phase
transitions swept across the Early Universe. The $^3$He-B neutron
experiment has been examined by solving analytically and
numerically the time-dependent Ginzburg-Landau equation
\cite{Aranson}. This study shows that in the non-equilibrium
transition of the neutron bubble there exists also a competing
surface mechanism, when the rapidly moving normal--superfluid
interface becomes unstable and vortex rings form around the bubble
in a corrugation instability. This is direct theoretical evidence
for the existence of the shear-flow instability also at the
normal--superfluid interface, between the surrounding superflow
and the stationary normal liquid within the bubble. A similar
situation is the normal--superfluid boundary in a gaseous rotating
Bose-Einstein condensate. Here the shear-flow instability also
leads to a deformation of the surface of the condensate cloud by
ripplons and to vortex-line formation \cite{VorBEC}.

The shear-flow instability thus appears to be a universal surface
mechanism for defect formation in superfluids and as intrinsic as
the KZ mechanism in the bulk liquid. We are left to wonder whether
or not these two are the only possible processes in the clean
limit at the lowest temperatures, to which other observations of
vortex formation can be reduced.




\begin{thebibliography}{99}
\bibitem{CritVelB} With only B phase in the sample, $v_{\rm cB}$ is determined
by single-vortex formation at the sample boundary or by the
leakage of vortices through the bottom orifice. Both processes
differ in appearance from the events in
Fig.~\protect\ref{InstabilitySteps}.
\bibitem{CritVelA} In this experiment the A-phase critical velocity
is $v_{\rm cA} \approx (0.36 \pm 0.06)$\,mm/s
\protect\cite{DoubleQuantumVortex}.
\end{thebibliography}
\end{document}